\documentclass[aps,twocolumn,superscriptaddress,showpacs]{revtex4}
\input epsf

\begin{document}

\title{Influence of mass polydispersity on dynamics of simple liquids
and colloids} 

\author{N. Kiriushcheva}
\email{nkiriusc@uwo.ca}
\affiliation{Department of Physics and Astronomy,
University of Western Ontario, London, Ontario N6A~3K7, Canada}

\author{Peter H. Poole}
\affiliation{Department of Applied Mathematics,
University of Western Ontario, London, Ontario N6A~5B7, Canada}

\date{\today }

\begin{abstract}
We conduct molecular dynamics computer simulations of a system of
Lennard-Jones particles, polydisperse in both size and mass, at a
fixed density and temperature. We test for and quantify systematic
changes in dynamical properties that result from polydispersity, by
measuring the pair distribution function, diffusion coefficient,
velocity autocorrelation function, and non-Gaussian parameter, as a
function of the degree of polydispersity. Our results elucidate the
interpretation of experimental studies of collective particle motion
in colloids, and we discuss the implications of polydispersity for
observations of dynamical heterogeneity, in both simulations of simple
liquids and colloid experiments.
\end{abstract}
\maketitle

\section{Introduction}

The dynamical behavior of liquids is an area of intense current
interest.  Much of this interest has been motivated by the desire to
understand the progressively slower and more complex dynamics of
dense, supercooled liquids as they are cooled toward the glass
transition~\cite{review}. In the last few decades, numerous direct
insights on dynamical motion in liquids have been obtained using
molecular dynamics (MD) computer simulations, in which the spatial
coordinates of particles as a function of time are
calculated~\cite{Allen}. More recently, experimental studies of
colloids have used confocal microscopy to track individual
particles~\cite{Kasper,Marcus,Weeks,Kegel}, thus generating the same
type of data on microscopic particle motions as is obtained from
simulations. For studying the glass transition, simulations and
colloid experiments therefore serve as important model systems in
which the implications of theory can be directly tested.

In both simulations and colloid experiments, fluids have been studied
in which the particle size is polydisperse. In simulations, size
polydispersity is often introduced to prevent crystallization of the
deeply supercooled liquid (see e.g.~\cite{KobAndersen}). In colloid
experiments, some degree of polydispersity is always present, and
depends on the process by which colloid particles are produced. To
characterize the polydispersity of colloids, the distribution $\theta
(\sigma )$ of particle diameters $\sigma $ is commonly found (or
assumed) to be Gaussian:
\begin{equation}
\label{thetaG}\theta _G(\sigma )=\frac 1{\delta \sqrt{2\pi }}\exp \left[
-\frac 12\left( \frac{\sigma -\sigma _0}\delta \right) ^2\right] , 
\end{equation}
where $\sigma _0$ is the average particle diameter, and $\delta$
characterizes the width of the
distribution~\cite{Hunter}. Polydispersity can then be quantified by
the value of the dimensionless parameter $c=\delta /\sigma _0$.

In most experimental studies of colloids (see e.g.~\cite{Hunter}) a
system is regarded as effectively monodisperse if $c<0.05$. For many
properties, such as the average liquid structure, this is a good
approximation. However, dynamical properties, especially at the
microscopic level, may depend sensitively on the nature of microscopic
structural fluctuations, and so may be affected by even small
polydispersities.  In addition, size polydispersity in real colloids
leads inevitably to a polydispersity of mass.  However, some models of
polydisperse liquids and colloids consist of systems in which particle
size varies, but not particle
mass~\cite{KobAndersen,DoliwaHeuer}.

In this paper we seek to isolate and quantify the role of size and
mass polydispersity on the dynamics of a simple liquid system, in
particular to assess the need to incorporate mass polydispersity when
simulating the dynamics of realistic systems having size
polydispersity.  To do so, we conduct MD simulations of a system of
particles interacting via the Lennard-Jones (LJ) potential,
polydisperse with respect to both mass and size, as a function of
$c$. Our results show that a range of dynamical properties (the
diffusion coefficient, the velocity autocorrelation function, and the
non-Gaussian parameter) of a polydisperse fluid are systematically
shifted from the corresponding monodisperse case. We discuss the
implications of these results for observations of ``dynamical
heterogeneity'' in simulations~\cite{DoliwaHeuer,kobdonati} and in
colloid experiments~\cite{Weeks,Kegel}.

\section{Polydisperse Lennard-Jones Liquid}

Since our aim is to study generic effects of polydispersity on liquid
dynamics, we chose the well-studied LJ potential to model
interparticle interactions.  The LJ potential is popular for
simulations of simple liquids, and there exist many studies to which
to compare our results.

In simulations of colloids, the colloidal particles are often modeled
as hard spheres, and for many cases this a good approximation.
However, interactions among colloidal particles can take other forms,
and can be explicitly controlled, for example, by attaching soluble
polymer chains by one end to the particle surface to generate
repulsion, or by adding nonadsorbing soluble polymers to the
suspension to produce attraction~\cite{GastRussel}.  Though the
present work is motivated by the recent experiments studying the
dynamics of colloidal particles, we do not address the question of how
the behavior of a colloidal system depends on the shape of the
interparticle interaction potential. We also do not take into account
the influence of a solvent.

We perform equilibrium molecular dynamics simulations in three
dimensions of a system of $N=4000$ particles interacting via the
shifted-force LJ potential, a modification of the standard LJ
potential,
\begin{equation}
\label{lj}V_{ij}(r)=4\varepsilon \left[ \left( \frac{\sigma _{ij}}r\right)
^{12}-\left( \frac{\sigma _{ij}}r\right) ^6\right] . 
\end{equation}
Here, $V_{ij}$ is the potential of interaction of two particles $i$
and $j$, separated by a distance $r$. $\varepsilon $ characterizes the
strength of the pair interaction and is constant for all particle
pairs. In the shifted-force LJ interaction, the LJ potential and force
are modified so as to go to zero continuously at $r=2.5\sigma _0$, and
interactions beyond $ 2.5\sigma _0$ are ignored~\cite{Allen}.

Polydispersity is introduced through the particle size:
$\sigma_{ij}=(\sigma_i+\sigma_j)/2$ where $\sigma_i$ ($\sigma_j$)
characterizes the diameter of a particle $i$ ($j$). Particles are
assigned $\sigma$ values by random sampling from the Gaussian
distribution in Eq.~\ref{thetaG}.  We also impose a mass
polydispersity appropriate for the given size polydispersity. The mass
of a particle $i$ is $m_i=m_0(\sigma_i/\sigma_0)^3$, where $m_0$ is
the mass of a particle of size $\sigma_0$.  Particle trajectories are
evaluated using the leap-frog Verlet algorithm~\cite{Allen}, using the
appropriate value of $m_i$ in the equation of motion of each particle.

Throughout this work we use reduced units. Energy is expressed in
units of $\varepsilon$, length in units of $\sigma_0$, the number
density of particles $\rho $ in units of $\sigma _0^{-3}$, and
temperature $T$ in units of $\varepsilon /k$, where $k$ is Boltzmann's
constant. Time $t$ is expressed in units of $\sqrt{m_0\sigma
_0^2/\varepsilon }$. In these units the time step used for integrating
the particle equations of motion is $0.01$.

After equilibration, all quantities are evaluated in the
microcanonical ensemble. We present data for $\rho =0.75$ and
$T=0.66$, a state not far from the triple point of the monodisperse LJ
fluid ($\rho =0.85$, $T=0.76$)~\cite{nic,BoonYip}. We chose this state
point so as to avoid the dense, deeply supercooled liquid region of
the phase diagram of the monodisperse LJ system, where spontaneous
crystallization could interfere with the evaluation of equilibrium
properties. We conduct separate simulations for $c=0$, $0.05$ and
$0.1$.

\section{Pair distribution function, diffusion coefficient, and velocity
autocorrelation function}

\begin{figure}
\hbox to\hsize{\epsfxsize=1.0\hsize\hfil\epsfbox{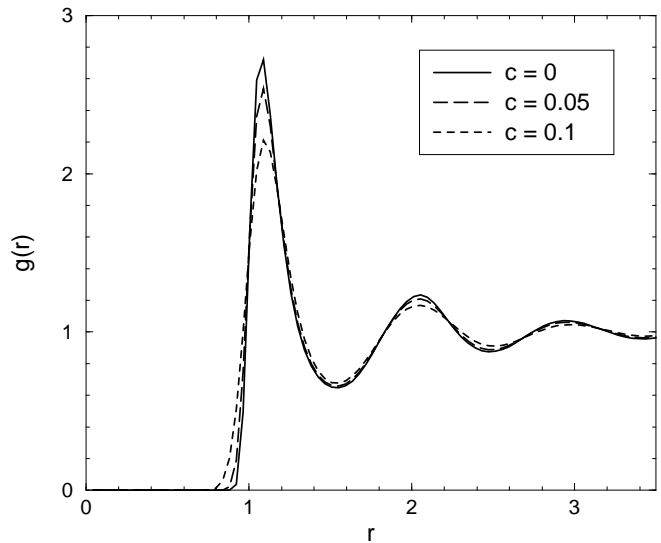}\hfil}
\caption{Effect of polydispersity $c$ on the average liquid structure
as measured by $g(r)$.}
\label{fig1}
\end{figure}

The pair distribution function $g(r)$ that characterizes the average
liquid structure~\cite{HansenMcDonald} is shown in Fig.~\ref{fig1} for
each $c$ studied. The effect of increasing polydispersity is to reduce
the height of, and broaden the peaks associated with the successive
neighbor shells around each particle. However the mean position of
each neighbor shell does not change noticeably.

\begin{figure}
\hbox to\hsize{\epsfxsize=1.0\hsize\hfil\epsfbox{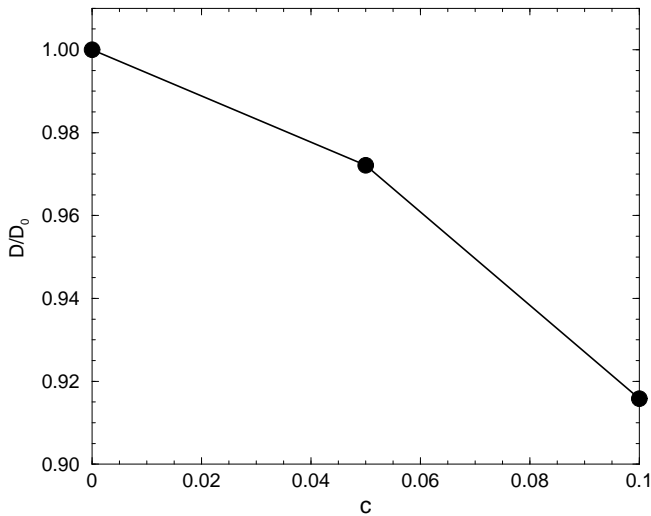}\hfil}
\caption{Fractional deviation with $c$ of $D$ relative to $D_0$, its
value for a perfectly monodisperse system.}
\label{diff}
\end{figure}

We test for a dependence on $c$ of the bulk transport properties by
evaluating the diffusion coefficient $D$. We obtain $D$ from $\langle
r^2(t)\rangle $ using the Einstein relation, 
\begin{equation}
D=\lim _{t\to \infty }\frac{\langle r^2(t)\rangle }{6t}. 
\end{equation}
Fig.~\ref{diff} shows the dependence of $D$ on $c$. We find that at
fixed $\rho $ and $T$, $D$ decreases systematically by about 10\% as
$c$ increases from zero to $0.1$.

\begin{figure}
\hbox to\hsize{\epsfxsize=1.0\hsize\hfil\epsfbox{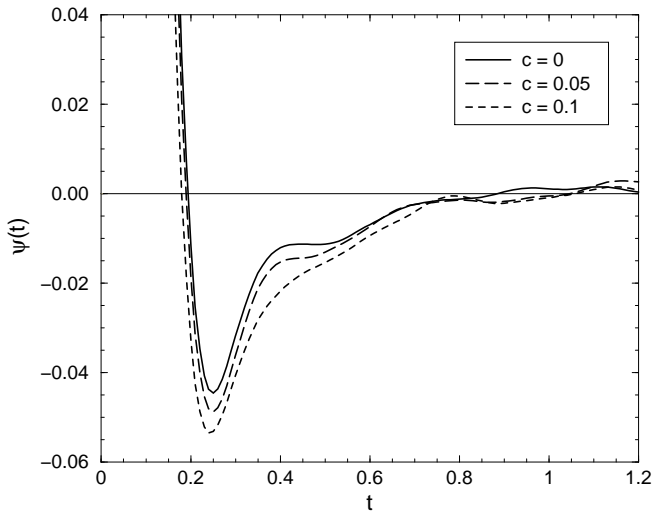}\hfil}
\caption{Effect of polydispersity $c$ on the velocity autocorrelation
function, $\psi(t)$.}
\label{vacf}
\end{figure}

Fig.~\ref{vacf} shows the dependence on $c$ of the velocity
autocorrelation function $\psi(t)$~\cite{BoonYip}:
\begin{equation}
\psi (t)=\frac{\left\langle {\bf v}(0)\cdot {\bf v}(t) 
\right\rangle} {\left\langle |{\bf v}|^2 \right\rangle} , 
\end{equation}
where ${\bf v}(t)$ is the velocity of a particle at time $t$.  $D$ is
related to the integral of $\psi(t)$, and consistent with the decrease
of $D$, the negative part of $\psi(t)$ becomes larger in magnitude
with increasing $c$. This trend reflects an increase with $c$ of the
strength with which single particle dynamical properties of the system
are retained on a time scale comparable to the collision time.

\section{Non-Gaussian parameter}

The general non-Gaussian parameter $\alpha _n(t)$ is defined for
integers $ n\geq 1$ as~\cite{BoonYip},
\begin{equation}
\label{alphan}\alpha _n(t)=\frac{\langle r^{2n}(t)\rangle }{c_n\langle
r^2(t)\rangle ^n}-1, 
\end{equation}
where $c_n=[{1\cdot 3\cdot 5\cdots (2n+1)}]/{3^n}$. $\langle
r^{2n}(t)\rangle $ is the ensemble average of the $2n^{{\rm th}}$
power of the particle displacements after a time $t$~\cite{Rahman}:
\begin{equation}
\label{rdef1}\langle r^{2n}(t)\rangle =\left\langle \frac 1N\sum_{i=1}^N|%
{\bf r}_i(t)-{\bf r}_i(0)|^{2n}\right\rangle . 
\end{equation}
Here, ${\bf r}_i(t)$ denotes the position of particle $i$ after a time $t$
following a reference time $t=0$ in equilibrium. $N$ is the total number of
particles in the system.

The functions $\langle r^{2n}(t)\rangle $ also represent the even
moments of $G_s({\bf r},t)$, the self-part of the van~Hove correlation
function~\cite{HansenMcDonald}.  For an isotropic fluid made up of
particles with spherically symmetric interactions, we can restrict our
attention to $G_s(r,t)$, the probability density that a particle
located at the origin at time $t=0$ will be found within $dr$ of a
distance $r$ at time $t$~\cite{vanHove}:
\begin{equation}
\label{Gs}G_s(r,t)=\left\langle \frac 1N\sum_{i=1}^N\delta (r-|{\bf r}_i(t)-%
{\bf r}_i(0)|)\right\rangle . 
\end{equation}
In terms of $G_s(r,t)$, $\langle r^{2n}(t)\rangle$ can be written,
\begin{equation}
\label{Gint}\langle r^{2n}(t)\rangle =4\pi \int_0^\infty
r^{2n}G_s(r,t)\,r^2dr . 
\end{equation}

In the case of an ideal gas of non-interacting particles having a
Maxwell-Boltzmann velocity distribution~\cite{HansenMcDonald},
$G_s(r,t)$ is a Gaussian function of $r$:
\begin{equation}
\label{Gauss}G_s(r,t)=\left( \frac{\beta m}{2\pi t^2}\right) ^{3/2}\exp
\left( -\frac{\beta mr^2}{2t^2}\right) ,
\end{equation}
where $\beta =1/kT$. In this case, it is readily shown that $\langle
r^{2n}(t)\rangle =c_n \langle r^2(t)\rangle ^n$ and so $\alpha
_n(t)=0$. For systems in which correlations of particle motions are
prominent, $G_s(r,t)$ is not Gaussian, and the deviation of $\alpha
_n(t)$ from zero serves to quantify the deviation of $G_s(r,t)$ from
the Gaussian form.

\begin{figure}
\hbox
to\hsize{\epsfxsize=1.0\hsize\hfil\epsfbox{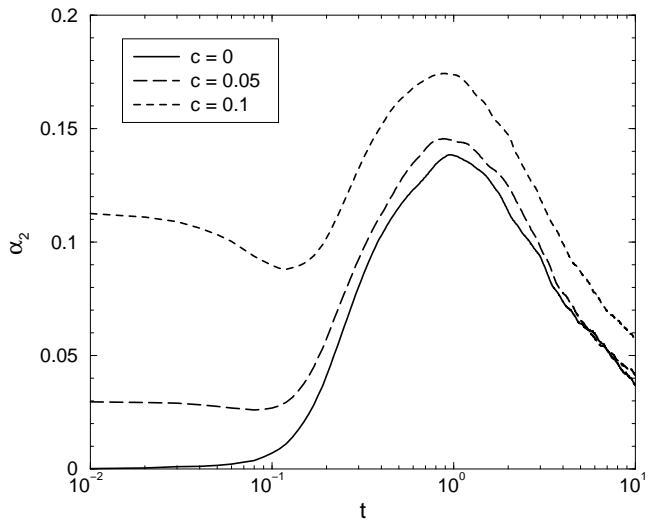}\hfil}
\caption{Variation of $\alpha_2(t)$ with $c$.}
\label{0510}
\end{figure}

In the present work we present results for $\alpha_2(t)$, the most
commonly calculated non-Gaussian parameter:
\begin{equation}
\label{alpha2}\alpha _2(t)=\frac{3\langle r^4(t)\rangle }{5\langle
r^2(t)\rangle ^2}-1.
\end{equation}
In Fig. \ref{0510} we plot $\alpha _2(t)$ for three different values
of polydispersity $c=0$, $0.05$ and $0.1$.  Qualitatively, there are
two effects induced by increasing polydispersity: (i) the
characteristic, intermediate-time peak of $\alpha _2$, at
approximately $t=1$, increases in magnitude as $c$ increases; and,
(ii) the value of $\alpha _2(t)$ does not start from zero in the limit
$t\rightarrow 0$ when $c\neq 0$.  We clarify each of these effects in
turn below.

\begin{figure}
\hbox to\hsize{\epsfxsize=1.0\hsize\hfil\epsfbox{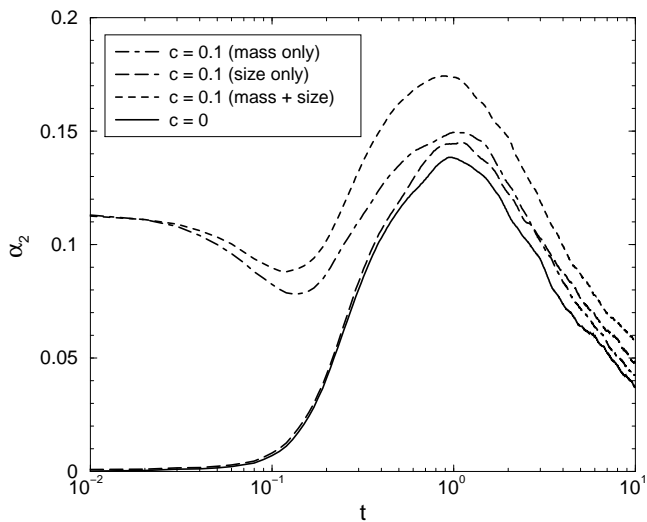}\hfil}
\caption{$\alpha_2(t)$ for several types of polydispersity,
demonstrating that polydispersity of both particle size and mass has
a greater impact, compared to the monodisperse case, than either
size-only or mass-only polydispersity.}
\label{alt}
\end{figure}

To distinguish the influence of mass and size polydispersity
separately, we conduct two new simulations, one (``size-only'') for a
system in which the size polydispersity is $c=0.1$, but in which all
the particle masses are set equal to $m_0$; and another
(``mass-only'') for which the mass polydispersity is taken from our
previous ``mass-and-size'' $c=0.1$ case, but with all the particle
sizes then set equal to $\sigma _0$. We compare in Fig.~\ref{alt} the
resulting behavior of $\alpha _2$ as a function of $t$ with the
behavior found for the monodisperse $c=0$ case; and with the case
where both size and mass are polydisperse with $c=0.1$.

\begin{figure}
\hbox to\hsize{\epsfxsize=1.0\hsize\hfil\epsfbox{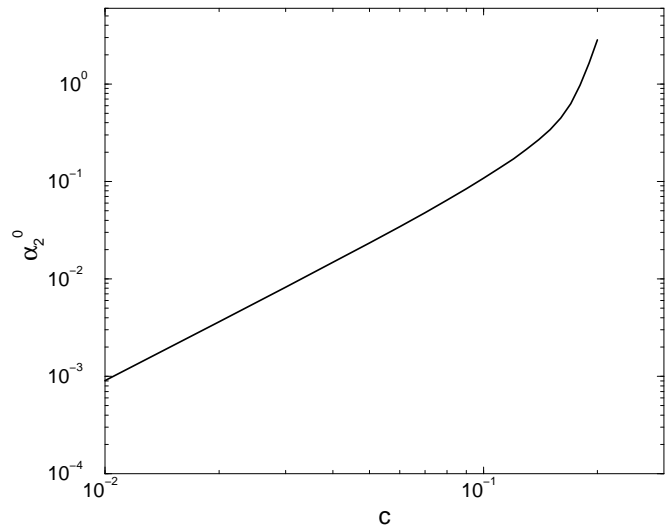}\hfil}
\caption{Log-log plot of $\alpha_2^\circ$ as a function of $c$ for a
Gaussian distribution of particle diameters.}
\label{alphac}
\end{figure}

First we focus on the behavior observed near the maximum of
$\alpha_2(t)$ at approximately $t=1$.  Although the mass-only curve in
Fig.~\ref{alt} lies above that of the monodisperse system, it still
lies well below that corresponding to polydispersity of both mass and
size. Hence, mass polydispersity is not solely responsible for the
increase of the maximum of $\alpha_2$ with $c$. Interestingly, the
size-only curve is also well below that corresponding to
polydispersity of both mass and size. Even the sum of the deviations
from the monodisperse case of the mass-only and size-only curves, is
insufficient to account for the height of the curve for the system
with polydispersity of both mass and size. That is, polydispersity of
both mass and size together has a greater impact on dynamical
properties at intermediate times than can be obtained from
polydispersity of mass or size alone.

Next we turn our attention to the behavior of $\alpha _2$ as $t\to
0$. (For the remainder of this work we will denote the limit as $t\to
0$ of $\alpha_2$ as $\alpha_2^{\circ}$.) In MD simulations of a
one-component LJ system~\cite{Rahman,Desai} and of a binary LJ
mixture~\cite{KobAndersen,kobdonati}, $\alpha _2^\circ=0$.  However,
in the binary LJ mixture studied, the two species differ in size only
and have the same mass, in contrast to our system in which the masses
of particles differ in accordance with the polydispersity of their
sizes.  Two of the curves in Fig.~\ref{alt} ($c=0$ and ``size-only'')
correspond to systems with no mass polydispersity, and in both cases
$\alpha_2^\circ=0$.  For the other two curves ($c=0.1$ and
``mass-only''), $\alpha_2^\circ$ adopts the same nonzero value. It is
clear that the mass polydispersity is solely responsible for the
behavior of $\alpha_2^\circ$.

As $t\to 0$, the atomic motions in the fluid correspond to those of
free particles, and the distribution of velocities is the Gaussian
function given in Eq.~\ref{Gauss}.  For a monodisperse system, this
means that $\alpha _2^\circ=0$.  For a system with polydisperse
masses, each particle of a given mass also samples the Gaussian
velocity distribution given in Eq.~\ref{Gauss}.  However, the Gaussian
distributions sampled will have different widths for particles of
different $m$.  Consequently, the form of the total $G_s(r,t)$
function is not in general Gaussian because it is a superposition of
individual Gaussians of different width.  The result is a non-zero
value of $\alpha_2^\circ$, as found in our simulations.

Since the limit $t\rightarrow 0$ corresponds to the free-particle
limit for atomic motion, we can calculate the non-Gaussian parameter
for a polydisperse system as $t\rightarrow 0$ exactly. Consider a
system of $N$ particles in which there are $M$ species (labeled by
index $j$) each having $N_j$ particles of mass $m_j$.  The moments
$\langle r^{2n}(t)\rangle$ can be found using a modified form of
Eq.~\ref{rdef1} appropriate for an $M$-component system:
\begin{equation}
\label{rdef2}
\langle r^{2n}(t)\rangle =
\left\langle 
\frac 1N\sum_{j=1}^M\sum_{i=1}^{N_j}
|{\bf r}_i(t)-{\bf r}_i(0)|^{2n} 
\right\rangle.
\end{equation}
Eq.~\ref{rdef2} can be rewritten as,
\begin{equation}
\label{rdefpar}\langle r^{2n}(t)\rangle 
=\sum_{j=1}^M f_j\langle r^{2n}(t)\rangle_j, 
\end{equation}
where $f_j=N_j/N$ is the fraction of particles of species $j$ and
\begin{equation}
\langle r^{2n}(t)\rangle_j = 
\left\langle 
\frac{1}{N_j} \sum_{i=1}^{N_j} |{\bf r}_i(t)-{\bf r}_i(0)|^{2n}
\right\rangle, 
\end{equation}
where the sum is over particles only of species $j$.  For each
species, the atomic motion as $t\to 0$ is also described by
Eq.~\ref{Gauss} with the appropriate value of $m=m_j$, and hence, the
moments $\langle r^2(t)\rangle _j$ and $\langle r^4(t)\rangle _j$ can
be found in the limit $t\to 0$ by substituting Eq.~\ref{Gauss} into
Eq.~\ref{Gint} for each $j$.  The result is,
\begin{equation}
\label{r2}\langle r^2(t)\rangle _j=\frac{3t^2}{\beta m_j} ,
\end{equation}
and,
\begin{equation}
\label{r4}\langle r^4(t)\rangle _j=\frac{15t^4}{\beta ^2m_j^2}. 
\end{equation}

The value of $\alpha _2^\circ$ for the multicomponent system can then
be found by using Eqs.~\ref{rdefpar}, \ref{r2} and \ref{r4} in
Eq. \ref{alpha2}:
\begin{equation}
\label{alpha0}
\alpha _2^{\circ }=\frac{ \sum_{j=1}^M m_j^{-2}f_j}{\left(
\sum_{j=1}^M m_j^{-1} f_j\right) ^2}-1
\end{equation}
This expression highlights that $\alpha_2^{\circ}$ may not equal zero
for a system with polydisperse masses.

If the polydispersity is expressed as a continuous distribution of
masses $\phi (m)$, Eq.~\ref{alpha0} generalizes to,
\begin{equation}
\label{genmass}\alpha _2^{\circ }=\frac{\int_0^\infty m^{-2}\phi (m)dm}{%
\left( \int_0^\infty m^{-1}\phi (m)dm\right) ^2}-1. 
\end{equation}

Experimental studies of colloids typically characterize polydispersity
not in terms of the masses, but in terms of the particle diameters,
described by $\theta (\sigma )$. Assuming that the particle mass $m$
is proportional to $\sigma ^3$, and that $\phi (m)\,dm=\theta (\sigma
)\,d\sigma $, Eq.~\ref{genmass} becomes,
\begin{equation}
\label{gensize}\alpha _2^{\circ }=\frac{\int_0^\infty {\sigma ^{-6}}\,{%
\theta (\sigma )}\,d\sigma }{\left( \int_0^\infty {\sigma ^{-3}}\,{\theta
(\sigma )}\,d\sigma \right) ^2}-1. 
\end{equation}
Note that the value of $\alpha _2^{\circ }$ therefore depends only on the
shape of the mass distribution function, and is otherwise constant for all
choices of $\rho $, $T$ and interparticle interaction.

We apply the above result to the case of the Gaussian distribution of
particle diameters given in Eq.~\ref{thetaG}. We substitute $\theta
=\theta _G$ with $\sigma _0=1$ in Eq.~\ref{gensize} and calculate
$\alpha _2^{\circ } $ as a function of $c$ (Fig.~\ref{alphac}). We
evaluate the integrals in Eq.~\ref{gensize} numerically, replacing the
limits of integration $(0,\infty )$ with
$[0.01\sigma_0,2\sigma_0]$. This avoids the divergence of the
integrands at $\sigma =0$, and in any case is more physical, since a
real distribution of particle sizes would have a non-zero lower bound,
and a finite upper bound. As seen in Fig.~\ref{alphac},
$\alpha_2^{\circ }\propto c^2$ for small $c$, but increases more
rapidly than this for $c>0.1$ The predictions of Eq.~\ref{alphac} are
in agreement with our simulation results.  For $c=0.05$
Eq.~\ref{gensize} gives $\alpha _2^{\circ }=0.02344$, while our
simulation gives $\alpha _2^{\circ }=0.027$; for $c=0.1$ we obtain
$\alpha _2^{\circ }=0.10781$ and $0.113$, respectively.

We can also use Eq.~\ref{gensize} to calculate $\alpha _2^{\circ}$ for
a system in which the distribution of sizes is not Gaussian. For
example, Sear~\cite{Sears} simulated a system of hard spheres,
polydisperse in both size and mass, using a ``hat'' function of width
$w$: $\theta (\sigma )=(w\sigma _0)^{-1}$ for $\sigma _0(1-w/2)\leq
\sigma \leq \sigma _0(1+w/2)$, and $\theta (\sigma )=0$ otherwise, and
$m\sim \sigma ^3$. For this case, we are able to solve
Eq.~\ref{gensize} exactly, giving,
\begin{equation}
\label{searseq}\alpha _2^{\circ }=\frac{(1+\omega /2)^5-(1-\omega /2)^5}{%
5\omega (1-\omega ^2/4)}-1. 
\end{equation}
For $\omega =0.3$, Eq.~\ref{searseq} gives $\alpha _2=0.06916$, while
for $\omega =0.7$, $\alpha _2=0.4222$, in agreement with the
simulation results in Fig.~6 of Ref.~\cite{Sears}.

\section{Discussion}

When $\alpha _2$ has been extracted via confocal microscopy in colloid
experiments, values as high as $\alpha _2^{\circ }\approx 1.5$ have
been observed~\cite{Weeks,Marcus,Kasper}. In these studies the
polydispersity ranged from $c=0.01$ to $0.1$. Kasper,
et~al.~\cite{Kasper} observed that $\alpha _2^\circ$ is not zero for
all values of the volume fraction occupied by the colloidal
particles. Marcus, et~al.~\cite{Marcus} found that $\alpha_2^{\circ }$
is nonzero and increases with increasing volume fraction.  Weeks, et
al.~\cite{Weeks} found that of $\alpha _2^\circ$ is approximately
constant for small volume fractions but grows for higher volume
fractions.  In general, $\alpha _2^{\circ}$ was found to increase with
the volume fraction occupied by the colloid particles, in contrast to
the absence of any density dependence in Eq.~\ref{gensize}. In the
case of real colloids, the behavior of $\alpha _2$ as $t\to 0$ is
complicated by the fact that solvent-induced hydrodynamic forces among
particles potentially introduce strong, short-time-scale correlations
of particle velocities, invalidating the free-particle assumption that
is the basis of Eq.~\ref{Gauss}. The large difference between the
behavior of $\alpha _2^{\circ }$ found for these systems, and that
predicted by Eq.~\ref{gensize} demonstrates that polydispersity alone
cannot account for the observed values of $\alpha _2^{\circ}$, and
that hydrodynamic effects indeed dominate the short-time dynamical
behavior of real colloids.

At intermediate times, we find that the peak value of $\alpha _2$
increases with $c$. The maximum value of $\alpha _2$ has been
shown~\cite{kobdonati} to correlate to the degree of dynamical
heterogeneity present in the system: that is, transient, spatially
correlated groups of particles whose characteristic structural
relaxation time differs from the mean.  Our results therefore suggest
that dynamical heterogeneity, prominently observed at lower $T$ and
higher $\rho$ than studied here, may be enhanced as polydispersity
increases. One source of this enhancement may be the influence of mass
polydispersity on spontaneously occurring density fluctuations, that
in turn control the development of dynamical heterogeneities.  In
general, the occurrence and quantification of dynamical heterogeneity
in a polydisperse system is likely to be more complicated than in a
monodisperse (or even bi-disperse) system. At the same time, since we
observe a slowing of the dynamics with increasing polydispersity, the
dynamical heterogeneity may be more prominent and longer-lived in a
polydisperse system, and so may facilitate the study of these complex
structures.

In summary, we have illustrated that a system of particles with
polydispersity of both mass and size is a more realistic model
(compared to models without mass polydispersity) for studying the
dynamics of colloids in MD simulations.  Our results show that typical
polydispersities found in real systems can induce an influence of the
order of 10\% on dynamical properties. This is a small effect for
studies, such as those near a glass transition, where relaxation times
may vary by several orders of magnitude.  At the same time, knowledge
of the amount and direction of the impact of polydispersity on
dynamics is required because polydispersity is so commonly found in
systems studied both in simulations and experiments. This knowledge is
also crucial for precise tests of theories, particularly those
formulated for perfectly monodisperse systems. We also note that our
results can be tested experimentally in colloids by deliberately
varying the polydispersity of the studied colloidal system. Indeed, by
varying $c$ alone it might be possible to study an
``isothermal-isochoric glass transition'' (i.e. a glass transition at
both fixed $T$ and $\rho $) by setting up an appropriate series of
colloidal systems where the polydispersity is progressively increased.

\section{Acknowledgments}

NK thanks Sergiy Kuzmin for valuable discussions. NK and PHP
acknowledge financial support from NSERC (Canada).

\end{document}